\documentclass{aa_new}
\usepackage{graphicx} 

\def\refe@jnl#1{{#1}}
\def\aj{\refe@jnl{AJ}}                   
\def\araa{\refe@jnl{ARA\&A}}             
\def\apj{\refe@jnl{ApJ}}                 
\def\apjl{\refe@jnl{ApJ}}                
\def\apjs{\refe@jnl{ApJS}}               
\def\ao{\refe@jnl{Appl.~Opt.}}           
\def\apss{\refe@jnl{Ap\&SS}}             
\def\aap{\refe@jnl{A\&A}}                
\def\aapr{\refe@jnl{A\&A~Rev.}}          
\def\aaps{\refe@jnl{A\&AS}}              
\def\aplett{\refe@jnl{Astrophys.~Lett.}} 
\def\apspr{\refe@jnl{Astrophys.~Space~Phys.~Res.}}
\def\azh{\refe@jnl{AZh}}                 
\def\baas{\refe@jnl{BAAS}}               
\def\bain{\refe@jnl{Bull.~Astron.~Inst.~Netherlands}}
\def\iaucirc{\refe@jnl{IAU~Circ.}}       
\def\fcp{\refe@jnl{Fund.~Cosmic~Phys.}}  
\def\gca{\refe@jnl{Geochim.~Cosmochim.~Acta}}   
\def\grl{\refe@jnl{Geophys.~Res.~Lett.}} 
\def\jcp{\refe@jnl{J.~Chem.~Phys.}}      
\def\jcap{\refe@jnl{J.~Cosm.~Astropart.~Phys.}} 
\def\jhep{\refe@jnl{J.~High.~Ener.~Phys.}}
\def\jgr{\refe@jnl{J.~Geophys.~Res.}}    
\def\jqsrt{\refe@jnl{J.~Quant.~Spec.~Radiat.~Transf.}}
\def\jrasc{\refe@jnl{JRASC}}             
\def\memras{\refe@jnl{MmRAS}}            
\def\memsai{\refe@jnl{Mem.~Soc.~Astron.~Italiana}}
\def\mnras{\refe@jnl{MNRAS}}             
\def\nat{\refe@jnl{Nature}}              
\def\npa{\refe@jnl{Nucl.~Phys.~A}}    
\def\npb{\refe@jnl{Nucl.~Phys.~B}}    
\def\plb{\refe@jnl{Phys.~Lett.~B}}       
\def\physrep{\refe@jnl{Phys.~Rep.}}      
\def\physscr{\refe@jnl{Phys.~Scr}}       
\def\pr{\refe@jnl{Phys.~Rev.}}           
\def\pra{\refe@jnl{Phys.~Rev.~A}}        
\def\prb{\refe@jnl{Phys.~Rev.~B}}        
\def\prc{\refe@jnl{Phys.~Rev.~C}}        
\def\prd{\refe@jnl{Phys.~Rev.~D}}        
\def\pre{\refe@jnl{Phys.~Rev.~E}}        
\def\prl{\refe@jnl{Phys.~Rev.~Lett.}}    
\def\planss{\refe@jnl{Planet.~Space~Sci.}}   
\def\procspie{\refe@jnl{Proc.~SPIE}}     
\def\pasp{\refe@jnl{PASP}}               
\def\pasj{\refe@jnl{PASJ}}               
\def\qjras{\refe@jnl{QJRAS}}             
\def\rmp{\refe@jnl{Rev.~Mod.~Phys.}}     
\def\rpp{\refe@jnl{Rep.~Prog.~Phys.}}    
\def\sci{\refe@jnl{Science}}             
\def\skytel{\refe@jnl{S\&T}}             
\def\solphys{\refe@jnl{Sol.~Phys.}}      
\def\sovast{\refe@jnl{Soviet~Ast.}}      
\def\ssr{\refe@jnl{Space~Sci.~Rev.}}     
\def\zap{\refe@jnl{ZAp}}                 

\newcommand{\BAR}{{\rm b}}

\newcommand{\ETAL}{{et al.\ }}
\newcommand{\EV}{{\rm eV}}
\newcommand{\MAT}{{\rm m}}

\newcommand{\MUK}{\mu{\rm K}}

\newcommand{\QUINT}{{\rm \scriptscriptstyle Q}}
\newcommand{\RMS}{{\rm rms}}
\newcommand{\SKY}{{\rm sky}}

\newcommand{\THEO}{{\rm theo}}

\def\gtrsim{\;\raise 0.4ex\hbox{$>$}\kern-0.7em\lower0.62ex\hbox{$\sim$}\;}
\def\lesssim{\;\raise 0.4ex\hbox{$<$}\kern-0.8em\lower0.62ex\hbox{$\sim$}\;}

\begin{document}

\title{An alternative to the cosmological `concordance model'}

\titlerunning{An alternative to the cosmological `concordance model}
\author{
A. Blanchard 
\inst{1}
\and
M.~Douspis 
\inst{2}
\and
M.~Rowan-Robinson
\inst{3}
\and
S.~Sarkar 
\inst{4}
}
\institute{
Laboratoire d'Astrophysique de l'Observatoire Midi-Pyr{\'e}n{\'e}es,
14 Avenue E.~Belin, F--31400 Toulouse, France
\and
Astrophysics, University of Oxford, Dennis Wilkinson Building, 
Keble Road, Oxford OX1 3RH, UK
\and
Astrophysics Group, Imperial College, Blackett Laboratory,
Prince Consort Road, London SW7 2BW, UK
\and
Theoretical Physics, University of Oxford, 1 Keble Road, Oxford OX1 3NP, UK}

\offprints{{\tt Alain.Blanchard@ast.obs-mip.fr}}

\date{\today}


\abstract{Precision measurements of the cosmic microwave background by
WMAP are believed to have established a flat $\Lambda$-dominated
universe, seeded by nearly scale-invariant adiabatic primordial
fluctuations.  However by relaxing the hypothesis that the fluctuation
spectrum can be described by a single power law, we demonstrate that
an Einstein-de Sitter universe with {\em zero} cosmological constant
can fit the data as well as the best concordance model. Moreover
unlike a $\Lambda$-dominated universe, such an universe has no strong
integrated Sachs-Wolfe effect, so is in better agreement with the low
quadrupole seen by WMAP. The main problem is that the Hubble constant
is required to be rather low: $H_0 \simeq 46$~km/s/Mpc; we discuss
whether this can be consistent with observations. Furthermore for
universes consisting only of baryons and cold dark matter, the
amplitude of matter fluctuations on cluster scales is too high, a
problem which seems generic. However, an additional small contribution
($\Omega_X \sim 0.1$) of matter which does not cluster on small
scales, e.g. relic neutrinos with mass of order eV or a `quintessence'
with $w \sim 0$, can alleviate this problem. Such models provide a
satisfying description of the power spectrum derived from the 2dF
galaxy redshift survey and from observations of the Ly-$\alpha$
forest.  We conclude that Einstein-de Sitter models can indeed
accommodate all data on the large scale structure of the Universe,
hence the Hubble diagram of distant Type Ia supernovae remains the
only {\em direct} evidence for a non-zero cosmological constant.  
{}}

\maketitle

\section{Introduction}

Measurements of cosmological parameters with reasonable accuracy are
essential both to establish a robust picture of the standard Big Bang
cosmology, and to provide insights into the fundamental processes, far
beyond the Standard Model of particle physics, which determined its
initial conditions. Since the pioneering work of Hubble, it has been
recognised that cosmological tests based on astrophysical arguments
can suffer from large systematic biases. Clearly one should as far as
possible use methods which do not depend explicitly on assumptions
concerning complex astrophysical phenomena.
 
In this respect, measurements of anisotopies in the Cosmic Microwave
Background (CMB) appear to offer the most promise for accurate
determination of cosmological parameters, thanks to the high control
possible on systematic errors. Since the epochal discovery of
primordial fluctuations on large angular scales by COBE (Smoot \ETAL
1992), this field has witnessed a renaissance. The
first detections of fluctuations on degree scales (Netterfield \ETAL
1995; Scott \ETAL 1996) provided tantalizing evidence for the flatness
of the Universe (e.g. Lineweaver \ETAL 1997). The unambiguous
detection of the first and second acoustic peaks in the angular power
spectrum (de Bernardis \ETAL 2000; Hanany \ETAL 2000; Halverson \ETAL
2002) has confirmed this result. Taken together with studies of
large-scale structure (LSS) in the universe, these observations have
also confirmed the overall picture of structure formation through
gravitational instability. The recent results obtained by WMAP
represent a further major advance in the field. For the first time,
measurements of cosmological parameters are being quoted with
uncertainties of a few per cent, opening up the anticipated era of
`precision cosmology'. Our intention here is to examine whether such
determinations are in fact robust or depend crucially on underlying
assumptions or `priors'. Specifically we wish to test whether a
cosmological constant, $\Lambda$, is really required by observations
of the CMB and LSS, {\em independently} of the indications from the
SN~Ia Hubble diagram. We will do so by confronting Einstein-de Sitter
(E-deS) models with the same observations. It turns out that with a
different assumption concerning the spectrum of primordial
fluctuations generated by inflation, such models can fit the data even
better than models with non-zero $\Lambda$.

\section{What do the $C_\ell$ measurements imply?}

The physics of passive linear perturbations in the early Universe is
well understood and therefore their evolution can be computed
accurately (see Hu \& Dodelson 2002). This is the basic reason why
precise measurements of the angular structure of the imprints left on
the CMB by primordial fluctuations can provide accurate information on
cosmological parameters. The ingredients necessary to compute the
amplitudes of the multipole moments ($C_\ell$) specifying the angular
power spectrum are both the nature and spectrum of the primordial
fluctuations (presumably arising from inflation), as well as
specification of the various contents of the universe which contribute
to its density and pressure.  The imprint of a specific parameter can
be direct, through the influence on the dynamics of acoustic
oscillations before the epoch of last scattering (as for the baryonic
content $\Omega_\BAR$ for instance), as well as indirect, through the
effect on the angular distance to the last scattering surface. A
non-zero $\Lambda$ affects the $C_\ell$s primarily through the
distance effect (Blanchard 1984). There are also more subtle effects,
such as the integrated Sachs-Wolfe (ISW) effect which contributes at a
much weaker level.  However such effects are harder to identify, as
they can easily be mimicked by a non-trivial primordial fluctuation
spectrum.

The first studies of the generation of density perturbations during
inflation established (see Linde 1990) that for the simplest models
involving a single `inflaton' field, the spectrum is close to the
Harrison-Zel'dovich (H-Z) scale-invariant form, $P(k) \propto k^n$
with $n = 1$, which had been proposed earlier on grounds of
simplicity. Thus the H-Z spectrum became a standard input for
calculations of CMB anisotropies and the growth of LSS, e.g. in the
standard cold dark matter (SCDM) model (Davis \ETAL 1985). In fact
there are significant corrections to a H-Z spectrum even in
single-field models, in particular the spectrum steepens
logarithmically with increasing $k$ (decreasing scale) as the end of
inflation is approached. This is usually accomodated by considering a
`tilted' spectrum with $n < 1$, although it should be noted that the
index $n$ is {\em scale-dependent} for any polynomial potential for
the inflaton, and is constant only for an exponential potential
(power-law inflation). Moreover $n$ can be close to, and even exceed,
unity if inflation ends not through the steepening of the inflaton
potential but, for example, due to the dynamics of a second scalar
field (hybrid inflation). In such multi-field models, the spectrum may
not even be scale-free since features can be imprinted onto the
spectrum, e.g. when the slow-roll evolution of the inflaton is
interrupted by other background fields undergoing symmetry-breaking
phase transitions (Adams, Ross \& Sarkar 1997b).

The expectations for the spectral index $n(k)$ in various inflationary
models has been reviewed by Lyth \& Riotto (1999). Even small
departures from scale-invariance can be quite significant for LSS
formation. For example after the SCDM model was found by Efstathiou
\ETAL (1992) to be in conflict with the observed power spectrum of
galaxy clustering (in being unable to simultaneously reproduce the
abundance of rich clusters (quantified by the variance $\sigma_8$ in a
top-hat sphere of radius $8h^{-1}$~Mpc) and the COBE measurement of
fluctuations on the scale of $H_0^{-1} \simeq 3000h^{-1}$~Mpc), it was
noted by White \ETAL (1995) that invoking a tilted spectrum with $n
\simeq 0.9$ could save the model. Interestingly enough, such a
spectrum arises from natural supergravity
inflation,\footnote{Technically `natural' means that the flatness of
the potential is protected by a symmetry --- here the shift symmetry
of a Nambu-Goldstone mode (Freese, Freiman \& Olinto 1990).} where the
leading term in the potential is {\em cubic} in the field (Ross \&
Sarkar 1996; Adams, Ross \& Sarkar 1997a). This yields $n =
(N-2)/(N+2)$, where $N \lesssim 57 + \ln~(k^{-1}/H_0^{-1})$ is the
number of e-folds of expansion from the end of inflation, taking the
inflationary energy scale to be $\lesssim10^{16}$~GeV as required by
the normalization to COBE, and the reheat temperature to be
$\lesssim10^{9}$~GeV to avoid the thermal gravitino problem (see
Sarkar 1996). If the inflationary scale is significantly lower
(German, Ross \& Sarkar 2001) and/or if there was a late epoch of
thermal inflation (Lyth \& Stewart 1996), then our present Hubble
radius may have exited the horizon only $\sim 20-30$ e-folds from the
end of inflation. This yields a spectral index as low as $n \simeq
0.8$ on cosmologically observable scales since $n \simeq 1 - 4/N$ in
this model.

It is thus clear that the primordial spectrum may not have a trivial
form and lacking a `standard model' of inflation, it is necessary to
consider a wide range of possibilities. Furthermore such complex
spectra could potentially confuse cosmological parameter estimation
from CMB data (e.g. Kinney 2001). This was explicitly demonstrated by
Barriga \ETAL (2001) using the COBE and BOOMERanG data (de Bernardis
\ETAL 2000) for the case of a primordial spectrum with a step-like
feature at a scale $k \sim 0.1~h/$Mpc, as in double inflation (Silk \&
Turner 1987). In this paper we investigate the flexibility in the
determination of cosmological parameters using the much more precise
WMAP data, when the usual hypothesis of a {\em single} power law
spectrum is relaxed. Since the power-law index is related to the slope
and curvature of the inflaton potential, it can change suddenly
e.g. if the mass of the inflaton changes through its coupling to a
background field which undergoes spontaneous symmetry breaking during
inflation (Adams \ETAL 1997b)

The possible detection of a non-zero $\Lambda$ through measurements of
the Hubble diagram of distant Type~Ia supernovae (Riess \ETAL 1998;
Perlmutter \ETAL 1999) is among the most significant developments in
modern cosmology (see Peebles and Ratra, 2003), and has led to the
establishment of the `concordance model' with $\Omega_\Lambda \sim
0.7$ and $\Omega_\MAT \sim 0.3$ (Bahcall \ETAL 1999). This is quite
consistent with the WMAP data for an assumed power-law primordial
spectrum; for a spatially flat universe the cosmological parameters
are determined to be: $\Omega_\MAT h^2 = 0.14 \pm 0.02$, $\Omega_\BAR
h^2 = 0.024 \pm 0.001$, $\Omega_\nu h^2< 0.0076$, $h=0.72 \pm 0.05$,
$n = 0.99 \pm 0.04$ and $\tau = 0.166^{+0.076}_{-0.071}$ (Spergel
\ETAL 2003). This agreement has led to to the widespread belief that
the $\Lambda$CDM concordance model is now established to high accuracy
through CMB measurements. However we wish to illustrate that this
agreement is crucially dependent on the underlying assumptions
concerning the primordial power spectrum and that CMB data do not yet
{\em independently} require a non-zero $\Lambda$. Before addressing
this issue, let us first assess the status of the concordance model.

\section{Is the concordance model actually concordant?}

\subsection{Concordance with WMAP data}

As mentioned already, SCDM with $\Omega_\MAT = 1$, $h = 0.5$ and $n =
1$ was found to disagree with the shape of the APM galaxy correlation
function (Maddox \ETAL 1990), as well as the high baryon fraction
measured in clusters, which together with the baryon fraction inferred
from primordial nucleosynthesis arguments, implied a lower matter
density of $\Omega_\MAT \sim 0.3$ in agreement with local dynamical
estimates (White \ETAL 1993). The subsequent measurements of CMB
fluctuations on degree scales however required the universe to be
spatially flat and ruled out such a low density matter-dominated
universe (Lineweaver \& Barbosa 1998). Thus the possible detection of
cosmic acceleration in the Hubble expansion of distant supernovae,
implying a cosmological constant with $\Omega_\Lambda \sim 0.7$, was
eagerly seized on as a mean of reconciling the CMB and LSS
data. However although this concordance model is consistent with most
cosmological observations, its first precision test has come with the
WMAP data. The agreement of the concordance model with data as
summarised by the WMAP team appears impressive (Spergel \ETAL
2003). However, there are two facts to keep in mind. First the global
$\chi^2$ on the temperature autocorrelation (TT) power spectrum is
rather poor --- the probability that the model fits the data is only
3\%, so strictly speaking the model is rejected at the $\sim 2\sigma$
confidence level!  However given possible remaining systematics
effects not yet accounted for, the WMAP team concluded that this
should not be considered as a serious problem for the concordance
model. Allowing for `running' of the spectral index with scale
improves the fit somewhat; the data suggest that $n \gtrsim 1$ on the
largest scales and $n \lesssim 1$ on small scales.

There is another aspect of the WMAP data that is even more puzzling,
viz. the amplitudes of the low $C_\ell$s, particularly the quadrupole,
is rather small compared to the expectation in the concordance model
where the large cosmological constant should boost the anisotropy on
large angles. It is well known that the cosmic variance is high on
such large angular scales and that Galactic foreground subtraction
introduces further uncertainties. Spergel \ETAL (2003) concluded, from
Monte Carlo realisations following two different methods, that the low
signal on large scales cannot be obtained in over 99\% of the cases.

However, statistical inferences from the quadrupole amplitude have to
be handled with caution. The measured value of $Q_{\rm rms} =
\sqrt{(5/4\pi)C_2} = 8 \pm 2~\MUK$ corresponds to a variance
($\Delta T_\ell^2 = \ell(\ell+1)C_\ell/2\pi$) of $\Delta T_2^2 =
154 \pm 70~\MUK^2$ (Bennett \ETAL 2003), so one might conclude that
e.g. an expected $\Delta T_2^2 = 350~\MUK^2$ is discrepant by about
$2\sigma$, which would be at the 95\% c.l. for a gaussian
distribution. In properly evaluating this probability however one
should take into account the foreground removal technique and its
uncertainty, in order to determine the likelihood distribution. Such a
distribution is likely to be non-Gaussian and it is therefore possible
that the estimation of the goodness-of-fit for the concordance model
might be significantly improved in this outlying region. If one
instead considers the best-fit concordance model derived by Spergel
\ETAL (2003) to be the true description of the CMB sky, the
probability of observing a low quadrupole can be directly
estimated. The log--likelihood of $C_\ell^\THEO$ can be well
approximated by (Bartlett \ETAL 2000):
$$
 -2 \ln \mathcal{L}(C_\ell^{\THEO}) = 
  5 \times f_{\SKY}\times \left[
  \ln \left(\frac{C_\ell + \aleph}{C_\ell^{\THEO} + \aleph}\right) 
  - \frac{C_\ell + \aleph}{C_\ell^{\THEO} + \aleph}\right]
$$
where the noise is $\aleph$ ($= 3.4 \MUK^2$ for $\ell = 2$ as quoted
in the WMAP data release\footnote{\tt http://lambda.gsfc.nasa.gov}),
the sky coverage is $f_\SKY = 85\%$, $C_\ell$ is the measured
amplitude and $C_\ell^\THEO$ is the value of the best-fit $\Lambda$CDM
model ($1204~\MUK^2$ for $\ell = 2$). Given this approximation, one
can retrieve the probability distribution of $C_2$ and thus estimate
the chance of observing a low value, following Douspis \ETAL
(2003). For $Q_\RMS = 8 \MUK$ this is 4.6\%, in agreement with the
estimate of Tegmark, de Oliveira-Costa \& Hamilton (2003) by a
different method (varying the cut CMB sky and considering the best
running spectral index model). This indicates that the quadrupole is
an outlier at most at the 2$\sigma$ level (see also Gazta\~naga \ETAL
2003, Efstathiou 2003b).

\subsection{Concordance with astronomical data} 


As discussed already, the concordance model has been built up over
time in order to match observations, thus its {\em a posteriori}
agreement with much of the LSS data is not a test. Of course as the
quality of data improves the model will be further tested, although
the number of free parameters provides some room for adjustment. 


Interestingly enough, WMAP has thrown new light on the masses of
galaxy clusters and therefore on the inferred baryon fraction which,
it had been argued (White \ETAL 1993), indicates a low matter density
universe. There has been some controversy in recent years concerning
the actual masses of X-ray emitting clusters, which are determined by
two different methods. One is the application of hydrostatic
equilibrium, while the second uses mass-temperature relationships
derived from numerical simulations. Systematic differences between the
two methods are significant (Markevitch 1998; Roussel, Sadat \&
Blanchard 2000) and this translates into an appreciable difference in
the derived value of $\sigma_8$ (Reiprich \& B\"ohringer 2002; Seljak
2002). Furthermore similar differences arise from the use of different
theoretical mass functions. The Sheth \& Tormen (1999) expression is
recognised as providing a satisfactory fit to the mass function
obtained from numerical simulations. Using this for an $\Omega_\MAT =
0.3$ universe, one finds $\sigma_8 = 0.86$ corresponding to the high
mass estimates from numerical simulations, and $\sigma_8 = 0.68$ for
the low mass estimates from hydrostatic equilibrium (Vauclair \ETAL
2003). Clearly the WMAP measurement of $\sigma_8 = 0.9 \pm 0.1$
(Spergel \ETAL 2003) favours the high mass estimates. The implied
baryon fraction (including stars) in clusters is then rather low,
slightly below 9\% (for $h = 0.7$). This {\em conflicts} with the
universal baryon fraction of 15\% required for the best concordance
model fit to the acoustic peaks! We note that this discrepancy
disappears for $\sigma_8 \sim 0.7$, which is about a $2\sigma$
deviation from the WMAP determination.

\section{Acceptable Einstein-de Sitter models}  

Let us now examine whether it is possible to obtain an acceptable CMB
power spectrum in an E-deS universe. Clearly to do this we have to
deviate from the assumptions of the $\Lambda$CDM concordance model
regarding the primordial power spectrum. Indeed the WMAP team (Peiris
\ETAL 2003) have already noted that the model fit can be significantly
improved (particularly to the outliers at $\ell =$ 22, 40 and 200) by
allowing for oscillations in the primordial spectrum (Adams, Cresswell
\& Easther 2001) such as might be induced by phase transitions
occuring during inflation (Adams, Ross \& Sarkar 1987b). However if we
are not to introduce too many new parameters, the simplest
modification that can be introduced is perhaps to consider a change in
the slope of the spectrum at a particular scale.  It is important in
this respect to notice that the first and second acoustic peaks span a
rather limited range of scales, $\ell \sim 150-600$, while the rising
part of the first peak covers a much bigger range, $\ell \sim
2-200$. Without advocating any specific scenario, it is interesting to
examine how a model with {\em different} power law indices for the
primordial fluctuations in these two regions compares to the
observational data. We therefore focus on models with 

$$ P(k) =
 \left\{ \begin{array}{ll} 
 A_1 k^{n_1} & {\rm for}\; k < k_1, \\ 
 A_2 k^{n_2} & {\rm for}\; k \geq k_1, \\ 
 \end{array}\right.  
$$ 
with a continuity condition ($ A_1 k_1^{n_1} = A_2 k_1^{n_2}$). We
calculate the CMB power spectrum using the CAMB code (Lewis \ETAL
2000) and use the WMAP likelihood code (Verde \ETAL 2003) to determine
the quality of the fit.

\begin{figure}[!ht] 
\begin{center} 
\resizebox{\hsize}{!}{\includegraphics[angle=0]{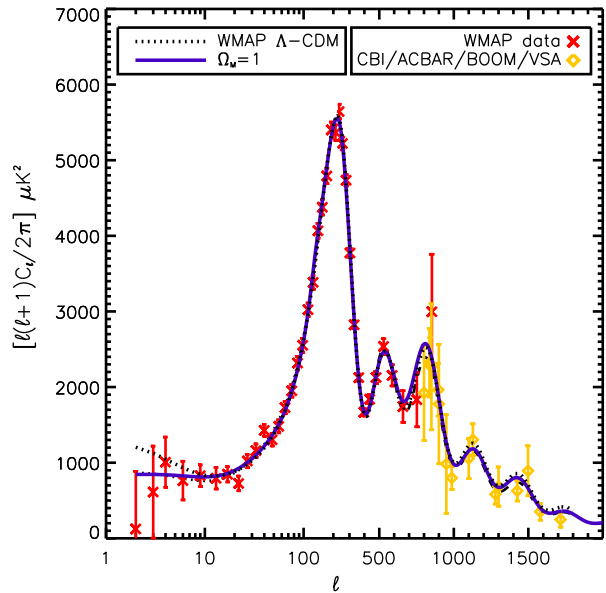}}
\end{center} 
\caption{\label{figOO} The temperature power spectrum for the best-fit
power-law $\Lambda$CDM model (dotted black line) from Spergel \ETAL
(2003), and for our broken-power-law model with $\Omega_\Lambda = 0$
(solid blue line), compared to data from WMAP and other experiments at
small scales --- CBI (\cite{cbi}), ACBAR (\cite{acbar}), BOOMERanG
(\cite{boom2}) and VSA (\cite{vsa}). Note the linear scale in $l$ for
$l > 200$. }
\end{figure}

As we are primarily interested in examining the possible constraints
on the cosmological constant, we have restricted our search to models
with $\Omega_\Lambda = 0$ but allow a reasonable range for other
cosmological parameters (including the optical depth $\tau$ to last
scattering). The best model we find has $h = 0.46$, $\omega_\BAR =
\Omega_\BAR h^2 = 0.019$, $\tau = 0.16$, $k_1 = 0.0096$~Mpc$^{-1}$,
$n_1 = 1.015$, $n_2 = 0.806$. As seen in Fig.~\ref{figOO} the
calculated power spectrum does very well in fitting the WMAP data and
other observations at high $\ell$. Interestingly enough a preferred
scale of $k \sim 0.01$~Mpc$^{-1}$ was also found by Mukherjee \& Wang
(2003) in attempting to reconstruct the primordial spectrum in the
context of a $\Lambda$CDM model (although Bridle \ETAL (2003) did not
detect this using a different method). We wish to emphasise that
inspite of having one additional parameter, our model has a better
$\chi^2$ (on the scalar $C_\ell$) than the best concordance model,
because of its lower amplitude at low $l$ (the $\chi^2$ of the TE
spectrum being identical).  In particular, the mean quadrupole $C_2$
has an amplitude of $844~\MUK^2$, which has a 13\% probability of
yielding $Q_\RMS = 8~\MUK$. The reason of course is that E-deS models
do not produce ISW effects as high as in flat models with low matter
density and a large cosmological constant. This is arguably a simpler
way to accomodate the observed low signal at $l \lesssim 20$, than to
invoke new physics (e.g. Spergel \ETAL 2003; Uzan, Kirchner \& Ellis
2003a; Efstathiou 2003a, Contaldi \ETAL 2003, Cline, Crotty \&
Lesgourgues 2003).

\subsection{Why this E-deS model is in difficulty} 

It might appear that our E-deS model with $\Omega_\MAT = 1$,
$\Omega_\Lambda = 0$ and $h = 0.46$ must be in conflict with a number
of astronomical observations. However, several of these observations
which in fact support the $\Lambda$CDM concordance model have been
questioned. For example, the observed mass-to-light function from
galaxies to superclusters yields $\Omega_\MAT = 0.16 \pm 0.05$
(Bahcall \ETAL 2000) but such observations are mostly local and
possibly untypical. The abundance of high redshift X-ray selected
clusters, a global test rather than a local one, systematically leads
to {\em high} values of $\Omega_\MAT$, well above the best WMAP value
(Henry 1997; Sadat, Blanchard \& Oukbir 1998; Viana \& Liddle 1999;
Borgani \ETAL 1999; Reichart \ETAL 1999; Blanchard \ETAL 2000). This
points in the same direction as the observed absence of any
large-scale correlations between the COBE map of the CMB and the
HEAO-1 map of the 2--8 keV X-ray background which provides an
interesting {\em upper limit} of $\Omega_\Lambda < 0.6$ at 95\%
c.l. (Boughn, Crittenden \& Koehrsen 2002).\footnote{Very recently,
Boughn \& Crittenden (2003) claim to have detected a large-scale
correlation between the WMAP and HEAO-1 maps at the $2.4-2.8\sigma$
level, which is surprising since the WMAP and COBE maps are quite
consistent. They also find a correlation between WMAP and the NVSS
survey of radio galaxies at the $1.8-2.3\sigma$ level. Nolta \ETAL
(2003) confirm the latter finding and reject a $\Lambda = 0$ universe
at 95\% c.l. However Myers \ETAL (2003) have found significant
contamination of the WMAP data by NVSS sources which can account for
the observed correlation. Fosalba \& Gazta{\~ n}aga (2003) also find a
correlation, twice as strong as the expected signal, between WMAP and
the APM galaxy survey.}

The only {\em direct} evidence so far for a cosmological constant
comes from the Hubble diagram of distant Type~Ia supernovae, a method
which relies on the standard candle hypothesis and on empirical
corrections to the observed peak magnitudes on the basis of the
observed decay times. Such corrections are essential for reducing the
scatter in the data sufficiently so as to allow significant
cosmological deductions. However there are systematic differences in
the corrections made for the {\em same} objects by the two groups
(Leibundgut 2000) which raises legitimate concerns about their
validity. Moreover, the distant SN~Ia appear to be significantly bluer
than the nearby sample, suggesting that the derived reddening may have
been underestimated (Leibundgut 2001). Rowan-Robinson (2002) has
argued that when extinction and the luminosity--decay time relation
are treated in a self-consistent way, the significance of the evidence
for positive $\Lambda$ is much reduced. 

\begin{figure}[!ht] 
\begin{center} 
\resizebox{\hsize}{!}{\includegraphics[angle=0]{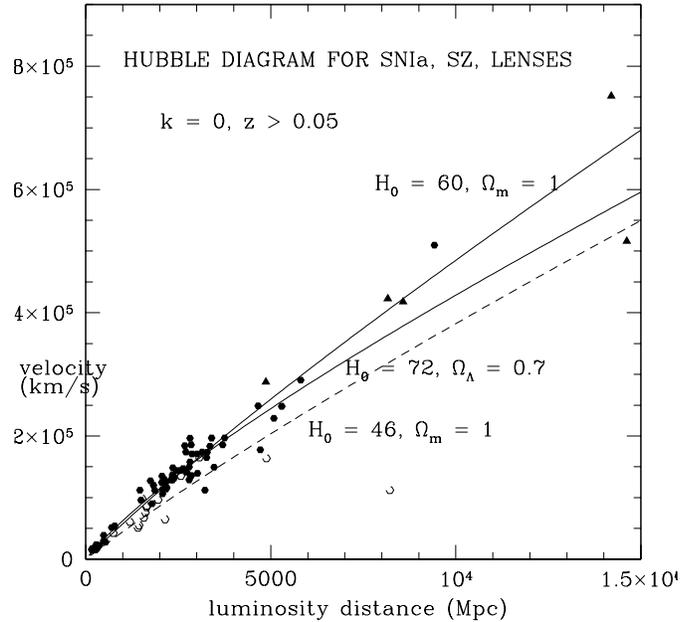}}
\end{center} 
\caption{\label{ho} Velocity versus luminosity-distance for Type Ia
supernovae (filled circles), S-Z clusters (open circles) and
gravitational lens time-delay systems (filled triangles), with z $>$
0.05. All curves shown correspond to flat models and are labelled with
the Hubble parameter in km/s/Mpc.}
\label{h0}
\end{figure}
 
A Hubble constant of $H_0 = 46$~km/s/Mpc would seem to be completely
inconsistent with the Hubble Key Project (HKP) determination of $72
\pm 8$~km/s/Mpc (Freedman \ETAL 2001).  However there are some details
of this work which might merit reexamination:

\begin{enumerate}

\item Rowan-Robinson (2003) finds that if a more sophisticated
local flow model is used than that of Mould \ETAL (2000), there is a
reduction of about 2~km/s/Mpc in $H_0$.

\item The method of combining the data used by Freedman \ETAL (2001),
viz. estimating $H_0$ by different methods and then combining the
results, does make the outcome somewhat vulnerable to Malmquist
bias. Rowan-Robinson (2003) finds that this can also result in
overestimation of $H_0$ by about 2~km/s/Mpc.

\item Between the summary paper of Mould \ETAL (2000) and that of
Freedman \ETAL (2001), a change in the assumed I-band
period-luminosity relation resulted in an increase of the extinction
values in the Cepheid program galaxies, and hence in an average
reduction in the distance scale of 4\%. This was almost exactly
cancelled out by correction for the effects of metallicity on Cepheid
distances, of about the same magnitude in the opposite
direction. However it is possible that these effects do not quite
cancel, e.g. if the extinction in the program galaxies has been
overestimated and/or the correction for the effects of metallicity
have been underestimated. It would be highly desirable to extend
observations of the Key Project Cepheids into the infrared to assess
these effects.

\item The assumed extinction in the LMC may be slightly on the low
side compared with estimates for hot stars in the LMC by Zaritsky
(1999).

\item Finally, it would be highly desirable to confirm the assumed
distance to the LMC, preferably by direct geometric methods. 

\end{enumerate}
All these effects are likely to be small, but it is possible that they
may combine in the same direction to significantly reduce $H_0$. 

It is interesting that methods which are largely {\em independent} of
the LMC and Cepheid distance scales, do tend to give significantly
lower values for $H_0$. For instance, Sunyaev-Zeldovitch (S-Z)
distances to 41 clusters give a value of $54^{+4}_{-3}$ km/s/Mpc in an
E-deS universe (Reese \ETAL 2002); furthermore any clumping of the
X-ray emitting gas would {\em lower} the actual value by up to
$\sim20\%$. The 4 simple gravitational lens systems (PG1115+080,
SBS1520+530, B1600+434, HE2149-2745) for which time delays have been
reliably measured yield $H_0 = 48 \pm 3$ km/s/Mpc if the lenses are
assumed to have isothermal halos of dark matter, while it would be $71
\pm 3$ km/s/Mpc if the lenses instead had constant $M/L$ (Kochanek \&
Schechter (2003). However Koopmans \ETAL (2003) have recently obtained
$H_0 = 75^{+7}_{-6}$ km/s/Mpc from a detailed reanalysis of the system
B1608+656, significantly higher than their previous estimate of $59
\pm 2$ km/s/Mpc for this system (Fassnacht \ETAL 2002). We note that
Parodi \ETAL (2000) find $H_0 = 59 \pm 4$ km/s/Mpc using SN~Ia (see
also Branch 2000).

Fig~\ref{h0} shows a compilation of distances to Type Ia supernovae,
S-Z clusters and gravitationally lensed systems with $z > 0.05$. The
best-fit flat model (not shown) has $\Omega_0 = 0.9 \pm 0.5$ and $H_0
= 60 \pm 11$ km/s/Mpc. Models with $\Omega_\MAT = 1$, $H_0 = 60$
km/s/Mpc and $\Omega_\MAT =0.3$, $\Omega_\Lambda = 0.7$, $H_0 = 72$
km/s/Mpc are shown (solid lines); both fit the data well. The model
with $\Omega_\MAT = 1$, $H_0 = 46$ km/s/Mpc (broken line) is clearly a
less good fit to the data. 

In conclusion a Hubble constant in the range 55-65 km/s/Mpc seems
entirely plausible at the present time. The value we require, 46
km/s/Mpc, is still below this range, but perhaps only by
$\sim1-2\sigma$. We believe the present paper provides a powerful
stimulus for further work on the cosmological distance scale.

\begin{figure}[!ht] 
\begin{center} 
\resizebox{\hsize}{!}{\includegraphics[angle=0]{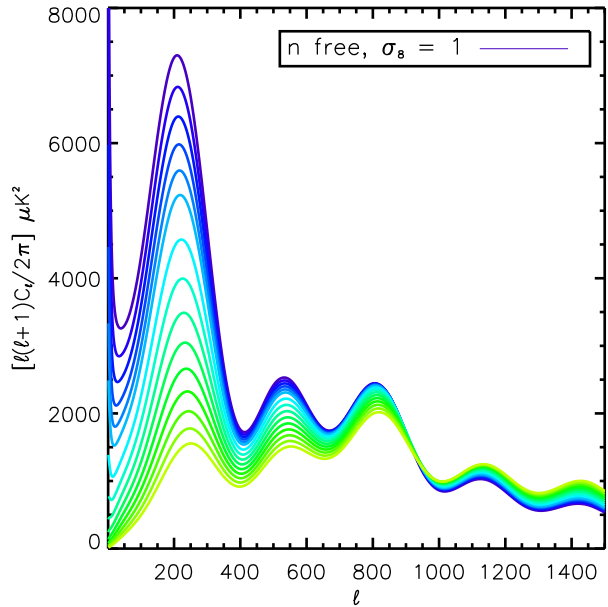}} 
\end{center} 
\caption{\label{sig8} CMB angular power spectra for models with
$\Omega_\MAT = 1$, $H_0 = 46$ km/s/Mpc and $\Omega_\BAR h^2 = 0.019$
normalised to $\sigma_8=1$ with different power law indices for the
primordial fluctuations. Note the constancy of the amplitude around
$\ell \sim 900$}
\end{figure} 

There is however one observational constraint that our E-deS model
fails to reproduce quite badly, viz. the amplitude of matter
fluctuations on cluster scales ($8h^{-1}$~Mpc). This model has an
amplitude of $\sigma_8 \sim 1.1$, which is much higher than required
to match the local abundance of clusters and weak lensing
measurements. The discrepancy is at least at the 5 $\sigma$ level,
even taking into account the scatter in the determinations from
different analyses: current estimates for $\sigma_8$ from clusters and
from weak lensing range from $0.45$ to $0.6$ (for $\Omega_\MAT
=1$). This is consistent with measurements of peculiar velocity
fields, e.g. the redshift-space distortion seen in 2dFGRS on scales
$<10h^{-1}$~Mpc yields $\beta \equiv \Omega_\MAT^{0.6}/b = 0.55 \pm
0.1$ for an E-deS universe, where $b \sim 1/\sigma_8$ is the linear
bias parameter (Hawkins \ETAL 2002).
\footnote{Their best-fit is $\beta = 0.49 \pm 0.09$ at the mean
redshift of the survey: $z \approx 0.17$. Combined with the estimate
of $b = 1.04 \pm 0.11$ from analysis of the bispectrum of 2dFGRS on
scales $k\sim (0.1-0.5)~h/$Mpc (Verde \ETAL 2002), this yields
$\Omega_\MAT (z=0) = 0.23 \pm 0.09$. However, this constraint on the
bias, obtained by an elaborate statistical analysis, may not be
reliable if the biasing process is more complicated than envisaged in
this work.} As WMAP cannot probe very small scales, one might imagine
that further modification of the primordial spectrum could remove this
discrepancy. For example Barriga \ETAL (2001) invoked a `step' in the
spectrum at $k\sim0.1$/h Mpc to decrease $\sigma_8$ significantly
below the corresponding value for a H-Z spectrum, and thus accounted
for the observed slow evolution of the number density of rich clusters
with redshift. However, an examination of the CMB power spectrum
normalized to $\sigma_8 = 1$ for various power law indices (see
Fig.~\ref{sig8}) reveals that the implied amplitude on the scale $l
\sim 900$ is essentially constant. Thus the recent measurements on
this scale by ground-based CMB experiments
(\cite{cbi,acbar,boom2,vsa}), if taken to be reliable, cannot be
matched if $\sigma_8$ is made significantly smaller than unity by
modifying the primordial spectrum. We conclude therefore that an E-deS
model with CDM {\em alone} cannot accommodate both data sets,
independently of the shape of the primordial spectrum.

\subsection{Modifying the matter content} 

In this last section, we examine whether the above discrepancy can be
avoided if we do not restrict ourselves to pure CDM models but modify
the matter content. Specifically, we now have evidence that all the
known neutrinos have masses which are rather close to each other, with
$\Delta\,m^2 \simeq 7\times10^{-5}~\EV^2$ for the electron and muon
neutrinos and $\Delta\,m^2\simeq3\times10^{-3}~\EV^2$ for the muon and
tau neutrinos, indicated respectively by the oscillation
interpretation of the Solar and atmospheric neutrino anomalies (see
Gonzalez-Garcia \& Nir 2003). Moreover the direct kinematic limit on
the neutrino mass from the Mainz and Troitsk tritium $\beta$-decay
experiments is 2.2~eV (see Weinheimer 2002). The addition of massive
neutrinos is known to damp the power spectrum on scales smaller than
their free-streaming length $d_\nu = 1230 (m_\nu/\EV)^{-1}$ Mpc, and
thus to lower $\sigma_8$ (see Primack \& Gross 2000). We have
therefore introduced 3 quasi-degenerate neutrinos of mass 0.8~eV each
and find the following model provides an acceptable fit: $h = 0.46$,
$\omega_\BAR = \Omega_\BAR h^2 = 0.021$, $\tau = 0.10$, $k_1 = 0.009$
Mpc$^{-1}$, $n_1 = 0.98$, $n_2 = 0.87$, $\Omega_\nu = 0.12$. The
amplitude on clusters scales is perhaps still too high at $\sigma_8 =
0.64$, but it is certainly premature to rule out this model for this
reason alone.

\begin{figure}[!ht] 
\begin{center} 
\resizebox{\hsize}{!}{\includegraphics[angle=0]{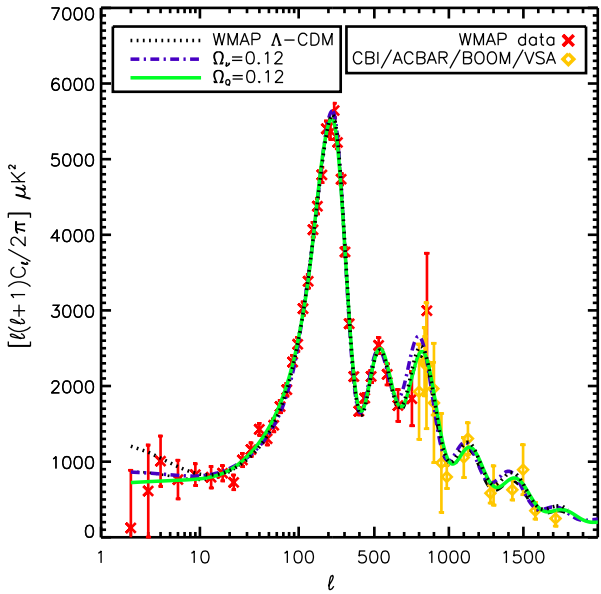}}
\end{center} 
\caption{\label{figNQ} The temperature power spectrum for the best-fit
power-law $\Lambda$CDM model (dotted black line) from Spergel \ETAL
(2003), and for our broken-power-law models (both having
$\Omega_\Lambda = 0$) with $\Omega_\nu = 0.12$ (dot-dashed blue line)
and $\Omega_\QUINT=0.12$ (solid green line), compared to data from
WMAP and other experiments (\cite{vsa,cbi,acbar,boom2}).}
\end{figure} 

In a second model we introduce a small amount of `quintessence' with
$w_\QUINT = 0$. Such possibility arises naturally as an attractor
solution of a simple exponential potential: $V = M_p^4 \exp (-\lambda
\Phi/M-p)$ which is well motivated theoretically 
(Ratra \& Peebles 1988; Wetterich 1988) and has been argued to 
give good agreement with
observations with $\Omega_\QUINT \sim 0.1$ (Ferreira \& Joyce 1997;
1998). Indeed, we find that the power at small scales is suppressed in
this model compared to pure cold dark matter, yielding an acceptable
fit to the CMB and LSS data (with $\sigma_8 = 0.5$) for the following
parameters: $h = 0.45$, $\omega_b = \Omega_b h^2 = 0.019$, $\tau =
0.10$, $k_1 = 0.012$~Mpc$^{-1}$, $n_1 = 1.00$, $ n_2 = 0.90$,
$\Omega_\QUINT=0.12$.

\begin{figure}[!ht] 
\begin{center} 
\resizebox{\hsize}{!}{\includegraphics[angle=0]{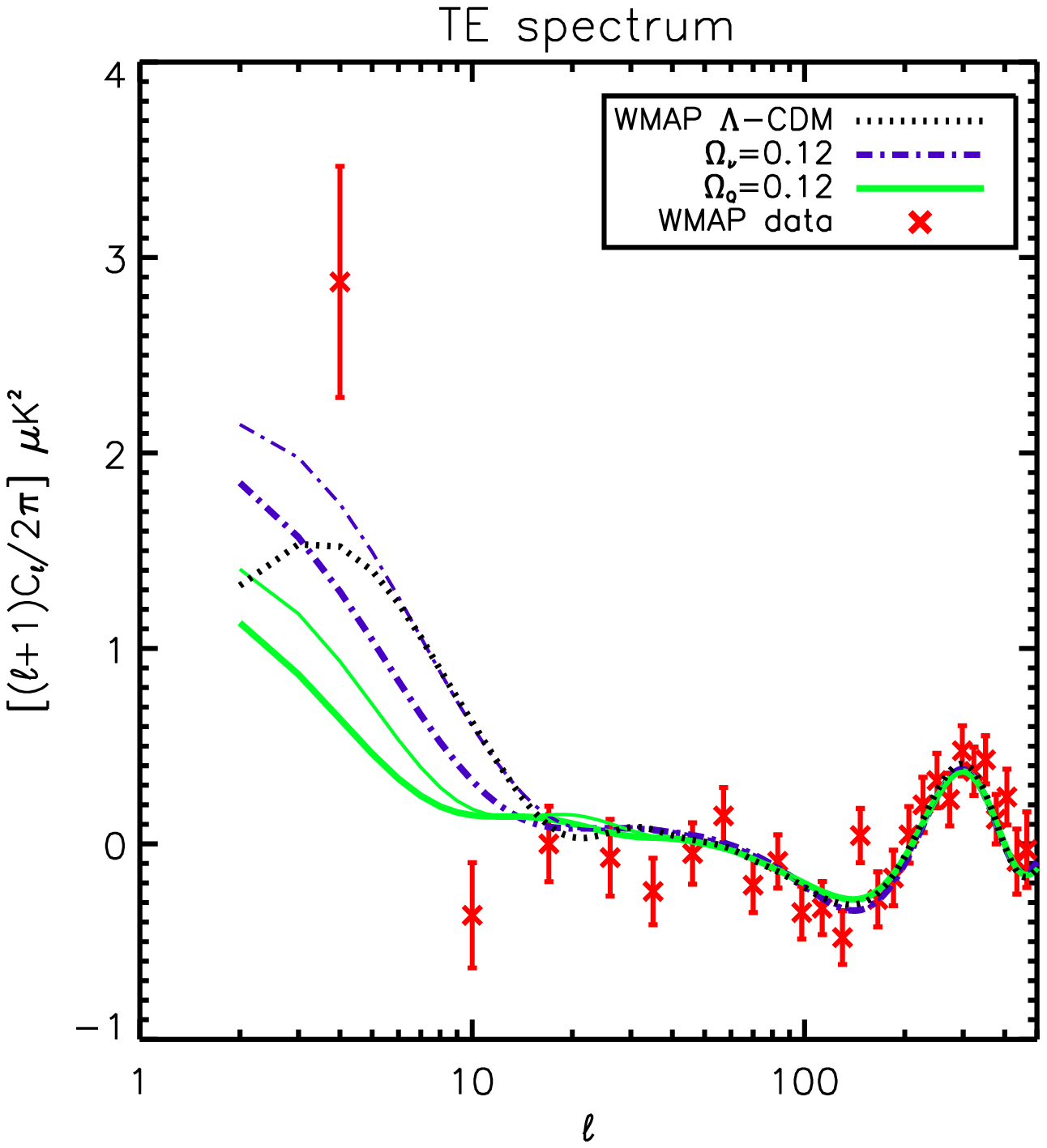}}
\resizebox{\hsize}{!}{\includegraphics[angle=0]{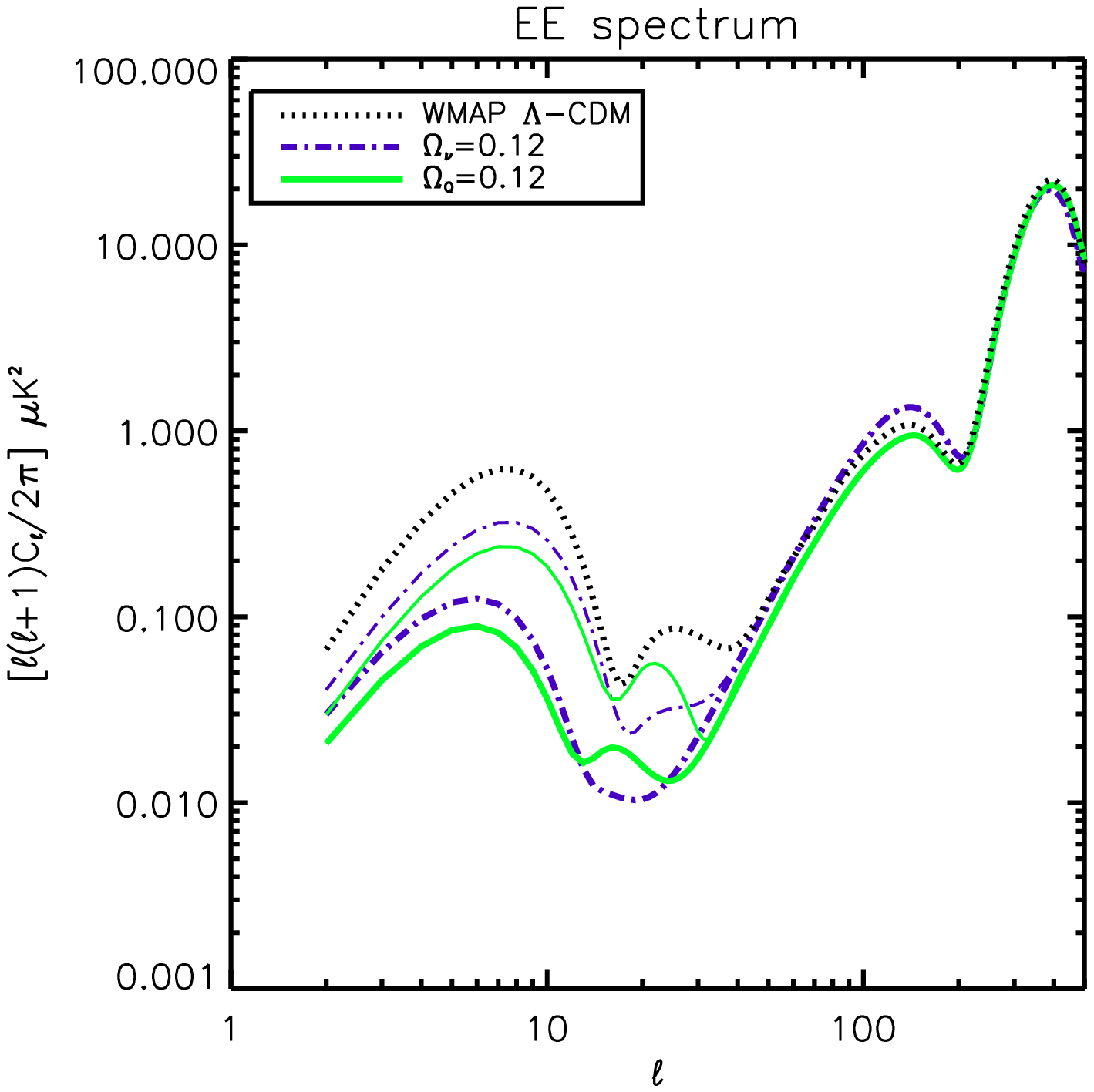}}
\end{center} 
\caption{\label{pol} The temperature-polarization (TE) cross
power-spectrum and the polarization (EE) power spectrum for our E-deS
models with $\Omega_\nu = 0.12$ (dot-dashed blue line) and
$\Omega_\QUINT=0.12$ (solid green line), both with $\tau = 0.1$,
compared to the concordance $\Lambda$CDM model (dotted line). The thin
lines are obtained adopting the higher optical depth ($\tau = 0.17$)
suggested by the WMAP fit to the TE data (Kogut \ETAL 2003).}
\end{figure} 

\begin{figure}[!ht] 
\begin{center} 
\resizebox{\hsize}{!}{\includegraphics[angle=0]{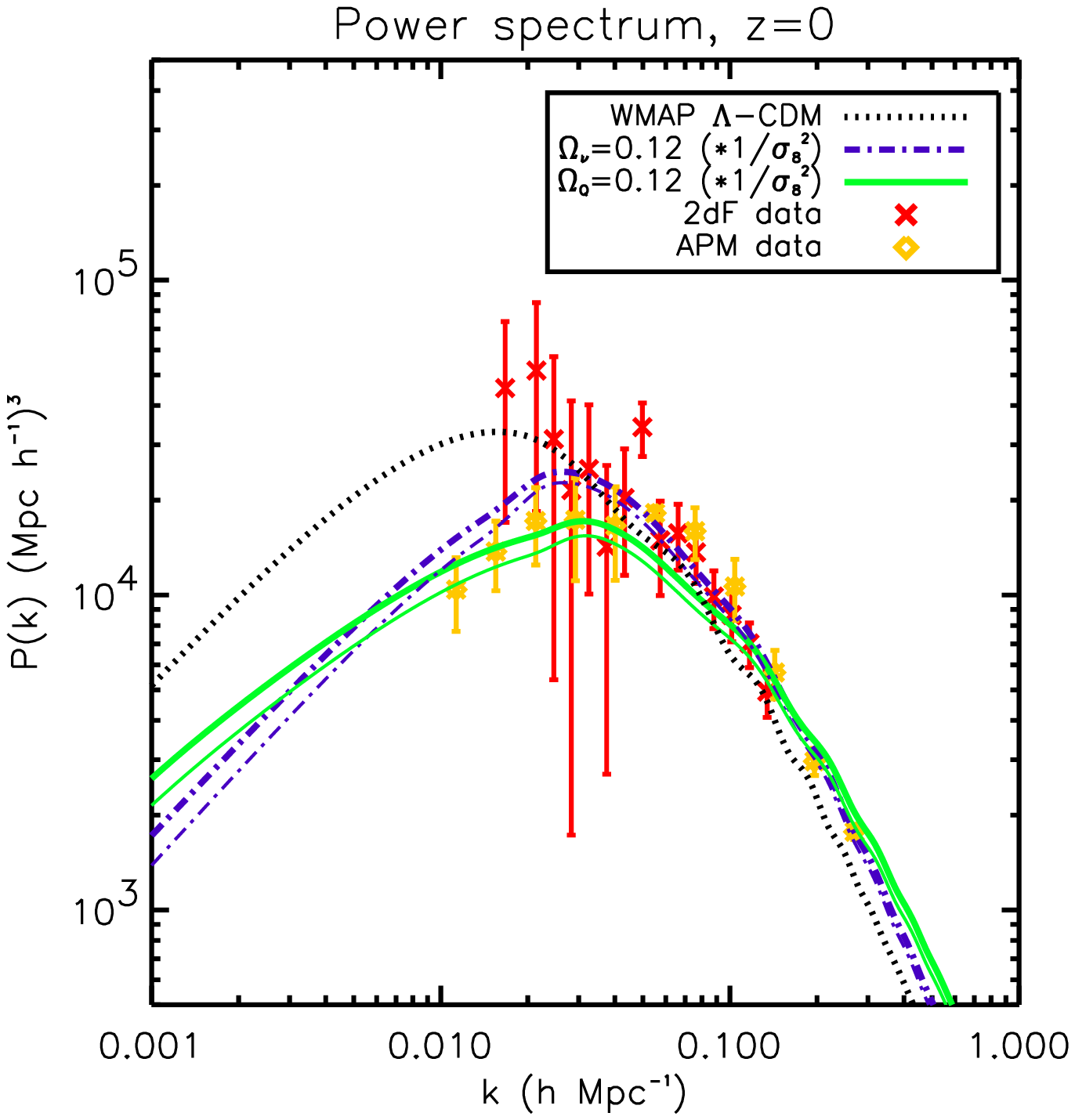}}
\resizebox{\hsize}{!}{\includegraphics[angle=0]{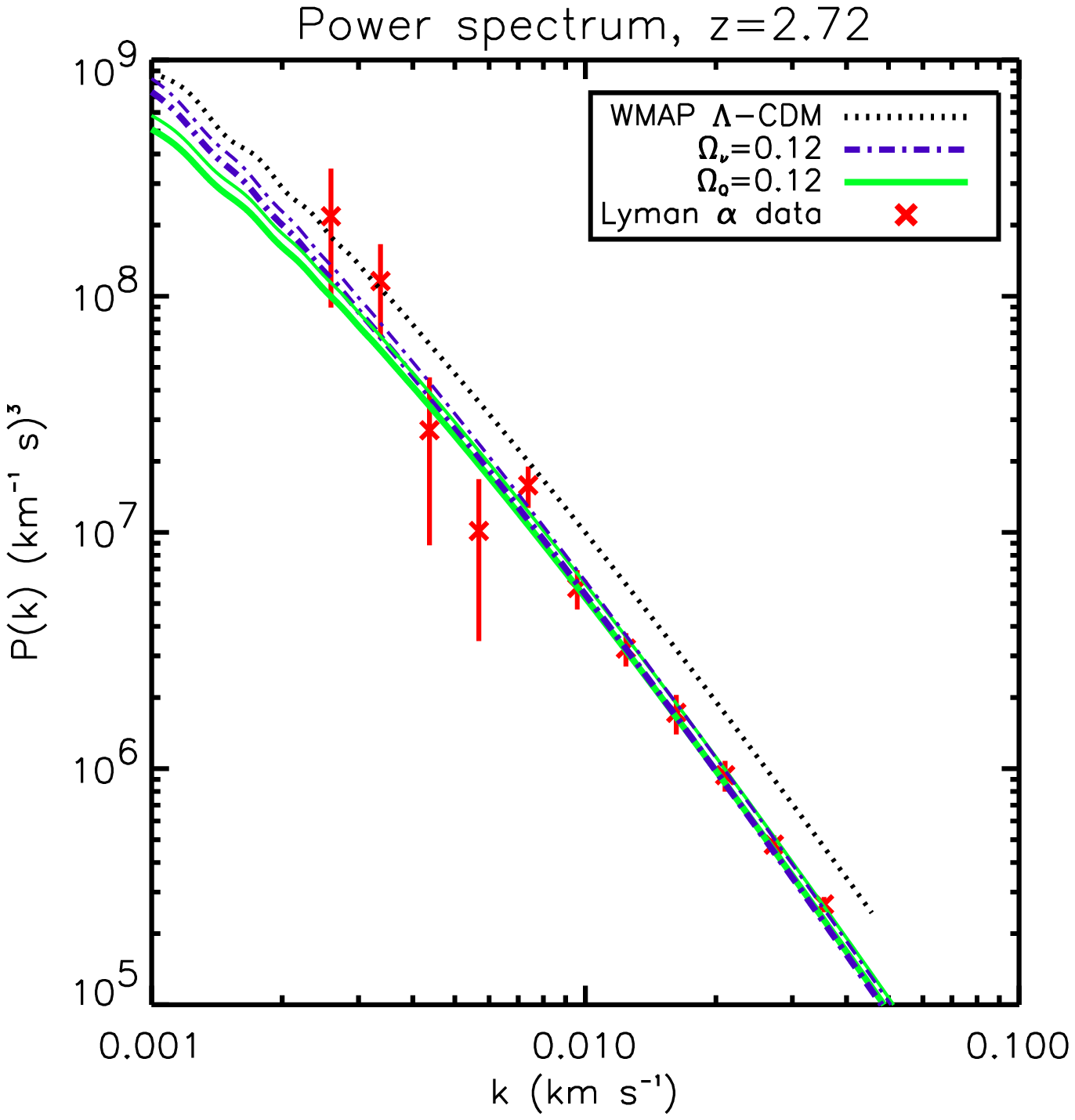}}
\end{center} 
\caption{\label{figPkNQ} The power spectrum of large scale structure
at $z=0$ and $z=2.3$ for our  E-deS models with $\Omega_\nu = 0.12$
(dot-dashed blue lines) and $\Omega_\QUINT=0.12$ (solid green
lines). The Lyman-$\alpha$ data have been shifted downwards by 20\%
(corresponding to the $1\sigma$ uncertainty in the calibration).}
\end{figure}

For both cases, the best-fit optical depth to last scattering $\tau
\simeq 0.1$ is significantly {\em smaller} than the value of $\tau =
0.17$ obtained in the WMAP team's fit to the temperature-polarization
cross-correlation (TE) spectrum (Kogut \ETAL 2003). Nevertheless the
predicted TE spectra for our E-deS models are in reasonable agreement
with the WMAP data, although the predicted polarization
autocorrelation (EE) spectra differ (see Fig.~\ref{pol}). The baryonic
content $\omega_\BAR = \Omega_\BAR h^2 = 0.019$ is at the upper end of
the range suggested by considerations of primordial nucleosynthesis
(Fields \& Sarkar 2002). The cluster baryon fraction expected in these two
models is $f_\BAR \sim 10-12\%$ (keeping in mind that the `dark
energy' component does not cluster on these scales); although low this
is consistent with recent estimates (Sadat \& Blanchard 2001). It is
also necessary to examine the agreement of these models with LSS data,
in particular the power spectrum obtained from studies of galaxy
clustering and the Lyman-$\alpha$ forest. For this purpose we adopt a
bias parameter given by $b = 1/\sigma_8$. As seen in
Fig.~\ref{figPkNQ}, both models are then in agreement with the APM
power spectrum (Baugh \& Efstathiou 1993; Peacock 1997), the 2dFGRS
power spectrum (Percival \ETAL 2001; Tegmark, Hamilton \& Xu 2002) and
the Ly$\alpha$ forest (matter) power spectrum (Croft \ETAL 2002) which
has been used recently to set interesting constraints on models of
large scale structure (Douspis, Blanchard \& Silk 2001). The model
with massive neutrinos provides a particularly good description of LSS
data. Such a model has already been considered by Elgar{\o}y \& Lahav
(2003) as providing a good fit to the 2dFGRS data, but maintaining a
{\em constant} power law index on large scales as they do, then gives
a very poor fit to the WMAP data. We emphasise that although our model
is apparently in conflict with the upper bound of $\Omega_\nu h^2 <
0.0076$ quoted by Spergel \ETAL (2003), this latter bound was obtained
under more restrictive asumptions (in particular adopting `priors' on
the bias parameter, matter density and Hubble parameter), hence is not
sufficiently conservative.

\section{Further tests}

It is important to devise means of breaking the
`$\Omega_\MAT-\sigma_8$ degeneracy' and establish reliably whether we
do live in an low matter density universe, and also to devise further tests
for $\Lambda$ independently of the SN~Ia Hubble diagram. 

Attempts to determine $\Omega_\MAT$ from measurements of peculiar
velocity fields have relied on constraining the bias parameter by
examination of the statistical properties of the galaxy distribution
(e.g. Verde \ETAL 2002). More detailed examination of the clustering
properties of different galaxy populations in the ongoing Sloan digital
sky survey will sharpen this test further (Szapudi \ETAL 2002).

Another approach is based on measurement of the mean relative peculiar
velocity of galaxy pairs as a function of their separation (Juskiewicz
\ETAL 2000); a recent application of this method yields $\Omega_\MAT =
0.29^{+0.15}_{-0.09}$, $\sigma_8 = 0.95^{+0.20}_{-0.11}$ without any
prior assumptions concerning the primordial fluctuations or
cosmological parameters (Feldman \ETAL 2003). However the inclusion of
a component of hot dark matter, as in our model, will affect peculiar
velocities on the relevant small scales, so simulations incorporating
non-linear effects are necessary to assess the robustness of this test
(J. Silk, private communication).

Measurements of `cosmic shear' induced through gravitational lensing
offer yet another possible way to determine the matter density and
bias separately (see Van Waerbeke \ETAL 2002). Again several ongoing
and proposed large area surveys should allow adequate control of
systematic uncertainties and enable cosmological parameters to be
obtained without prior assumptions.

A complementary approach is to seek direct evidence in the CMB for the
presence of a cosmological constant. As mentioned earlier, the absence
of the expected ISW effect at large angular scales in the WMAP data
has been disappointing in this respect, but might be regarded as a
statistical fluctuation. The expected ISW correlations with other
tracers of large-scale structure are being sought but results are not
definitive as yet. The power spectrum of the E-mode polarization
offers an additional way to distinguish among models. As seen in
Fig.~\ref{pol}, the nominal EE spectra for our E-deS models differ
significantly from that of the concordance model since the best-fit
value of $\tau$ is smaller. However the value of $\tau$ for our E-deS
models can be raised to be closer to the WMAP value of 0.17 (thin
lines in Fig.~\ref{pol}) without significantly affecting the $C_l$ or
$P(k)$ fits. Hopefully analysis of further data from WMAP, as well as
other CMB experiments, can distinguish between these possibilities.
The most stable difference between our E-deS models and the
$\Lambda$CDM concordance model is in fact the matter power spectrum
shape in the range $k \sim (0.01-0.03)~h$/Mpc, which galaxy surveys
may be able to investigate, provided the possible biasing is reliably
understood on these scales (Durrer \ETAL 2003).

\section{Discussion}  

We have shown that when the assumption of a {\em single} power law for
the primordial fluctuation spectrum is relaxed, an Einstein-de Sitter
model with zero $\Lambda$ can fit the CMB data as well as if not
better than the best $\Lambda$CDM concordance model. This is a clear
and direct indication that the CMB data alone does not require the
introduction of a non-zero cosmological constant. However a model with
only cold dark matter cannot simultaneously match both the CMB data
and the amplitude of matter fluctuations as indicated by clusters,
peculiar velocity fields and weak lensing measurements. We have shown
further that acceptable Einstein-de Sitter models are indeed possible
provided they comprise a small amount of a `dark energy' component
which does not cluster on small scales, such as relic neutrinos with
masses of order eV or a pressureless quintessence field. These models
have a low, but as we have argued, not unimaginable, Hubble constant.
Moreover this provides further motivation for laboratory experiments
sensitive to eV-scale neutrino masses, since a detection would provide
crucial input for cosmology.

Given the need to suppress the amplitude of fluctuations on cluster
scales in {\em any} model, we conclude that extant CMB and LSS data
actually imply the existence of a dark component beyond cold dark
matter, with a density contribution of about 10\% of the critical
density and an equation of state corresponding to conventional
pressureless matter. Such models reproduce quite well the observed
properties of the large scale structure of of universe without further
adjustment.

The essential conclusion is that an Einstein-de Sitter universe is
{\em not} yet ruled out, as seems to be generally believed. Although
there is indeed conflict with some astronomical observations, we have
argued that these data are not established beyond reasonable
doubt. Given the severe coincidence problem associated with a
classical cosmological constant as well as the profound implications
of de Sitter space-time for fundamental physics (e.g. Witten 2001,
Banks \& Dine 2001, Dyson, Kleban \& Susskind 2002), it
is surely worth investigating these issues further.

\begin{acknowledgements}  

We thank the Referee for her critical and helpful comments. We also
acknowledge useful discussions with David Spergel, Alain Riazuelo, Joe
Silk and Ludovic Van Waerbeke. M.D. is supported by a CMBNet
fellowship.
 
\end{acknowledgements}


\begin{thebibliography}{99}  

\bibitem{Adams01}
Adams, J.~A., Cresswell, B., \& Easther, R.\ 2001,
\prd, 64, 123514

\bibitem{Adams97a}
Adams, J.~A., Ross, G.~G., \& Sarkar, S.\ 1997a,
\plb, 391, 271 

\bibitem{Adams97b}
Adams, J.~A., Ross, G.~G., \& Sarkar, S.\ 1997b, 
\npb, 503, 405

\bibitem[Bahcall \ETAL 2000]{M/L} 
Bahcall, N.~A., Cen, R., Dav\'e, R., Ostriker, J.~P., \& Yu, Q.\ 2000, 
\apj, 541, 1

\bibitem[Bahcall \ETAL 1999]{triangle} 
Bahcall, N.~A., Ostriker, J.~P., Perlmutter, S., \& Steinhardt, P.~J.\
 1999,
\sci, 284, 1481

\bibitem{bd}
Banks, T., \& Dine, M.\ 2001,
\jhep, 0110, 012

\bibitem[Barriga \ETAL 2001]{barriga} 
Barriga, J., Gazta{\~ n}aga, E., Santos, M.~G., \& Sarkar, S.\ 2001, 
\mnras, 324, 977

\bibitem[Bartlett \ETAL 2000]{bartlett} 
Bartlett, J.~G., Douspis, M., Blanchard, A., \& Le Dour, M.\ 2000, 
\aaps, 146, 507 

\bibitem[Baugh \& Efstathiou 1993]{baugh} 
Baugh, C.~M., \& Efstathiou, G.\ 1993, 
\mnras, 265, 145 

\bibitem[Bennett \ETAL 2003]{bennett} 
Bennet C.L. \ETAL (WMAP collab.)\ 2003, 
\apj, submitted {\tt [astro-ph/0302207]}

\bibitem{blan}
Blanchard, A.\ 1984,
\aap, 132, 359

\bibitem[Blanchard \ETAL 2000]{blanchard00}  
Blanchard, A., Sadat, R., Bartlett, J.~G., \& Le Dour, M.\ 2000, 
\aap, 362, 809  

\bibitem[Borgani \ETAL 1999]{borgani}  
Borgani, S., Rosati, P., Tozzi, P., \& Norman, C.\ 1999, 
\apj, 517, 40 

\bibitem{bc}
Boughn, S.~P., \& Crittenden, R.\ 2003,
\nat, submitted {\tt [astro-ph/0305001]}

\bibitem[Boughn, Crittenden \& Koehrsen (2002)]{boughn}
Boughn, S.~P., Crittenden, R.~G., \& Koehrsen, G.~P.\ 2002,
\apj, 580, 672

\bibitem{branch} 
Branch, D.\ 1998
\araa, 36, 17

\bibitem{blwe03}
Bridle, S., Lewis, A.M., Weller, J., \& Efstathiou, G.\ 2003,
\mnras, submitted {\tt [astro-ph/0302306]}
              
\bibitem{ccl}
Cline, J.M., Crotty, P., \& Lesgourgues, J.\ 2003,
\jhep, submitted {\tt [astro-ph/0304558]}

\bibitem{cpkl}
Contaldi, C., Peloso, M., Kofman, L., \& Linde, A.\ 2003,
\prd, submitted  {\tt [astro-ph/0303636]}

\bibitem[Croft \ETAL 2002]{lya} 
Croft, R.~A.~C., Weinberg, D.~H., Bolte, M., Burles, S., Hernquist, L., 
 Katz, N., Kirkman, D., \& Tytler, D.\ 2002, 
\apj, 581, 20

\bibitem{cdm}
Davis, M., Efstathiou, G., Frenk, C.~S., \& White, S.~D.~M.\ 1985
\apj, 292, 371

\bibitem{boom} 
de Bernardis, P. \ETAL (BOOMERanG collab.)\ 2000,
\nat, 404, 955

\bibitem[Douspis \ETAL (2003)]{gof} 
Douspis, M., Bartlett, J.~G., \& Blanchard, A.\ 2003, 
\aap, submitted

\bibitem[Douspis, Blanchard \& Silk 2001]{dbs} 
Douspis, M., Blanchard, A., \& Silk, J.\ 2001, 
\aap, 380, 1 

\bibitem[Durrer, Gabrielli, Joyce, \& Sylos Labini(2003)]{2003ApJ...585L...1D} 
Durrer, R., Gabrielli, A., Joyce, M., \& Sylos Labini, F.\ 2003, 
\apjl, 585, L1 

\bibitem{dks}
Dyson, L., Kleban, M., \& Susskind, L.\ 2002
\jhep, 0210, 011

\bibitem[Efstathiou 2003a]{efst03a} 
Efstathiou G.\ 2003a, 
\mnras, in press {\tt [astro-ph/0303127]}

\bibitem[Efstathiou 2003b]{efst03b} 
Efstathiou G.\ 2003b, 
\mnras, submitted {\tt [astro-ph/0306431]}

\bibitem[Efstathiou, Bond \& White (1992)]{ebw92} 
Efstathiou, G., Bond, J.~R., \& White, S.~D.~M.\ 1992, 
\mnras, 258, 1P 

\bibitem[Elgaroy \& Lahav 2003]{lahav} 
Elgar{\o}y O., \& Lahav O.\ 2003,
\jcap, 0304, 004 

\bibitem{lensing} 
Fassnacht, C.~D., Xanthopoulos, E., Koopmans, L.~V.~E., \& Rusin, D.\ 2002,
\apj, 581, 823

\bibitem{feld}
Feldman, H.~A. \ETAL\ 2003,
{\tt [astro-ph/0305078]}

\bibitem{fs02}
Fields, B., \& Sarkar, S.\ 2002,
\prd, D66, 010001-162

\bibitem{fg97}
Ferreira, P.~G. \& Joyce, M.\ 1997, 
\prl 79, 4740
 
\bibitem{fg98}
Ferreira, P.~G. \& Joyce, M.\ 1998, 
\prd 68, 3503
 
\bibitem{fg03}
Fosalba, P., \& Gazta{\~ n}aga, E.\ 2003,
\prl, submitted {\tt [astro-ph/0305468]}

\bibitem[Freedman \ETAL 2001]{freedman} 
Freedman, W.~L. \ETAL (HST Key Project collab.)\ 2001, 
\apj, 553, 47

\bibitem[Freese, Frieman, \& Olinto (1990)]{ffo90} 
Freese, K., Frieman, J.~A., \& Olinto, A.~V.\ 1990, 
\prl 65, 3233 

\bibitem{loQ}
Gazta\~naga, E., Wagg, J., Multam\"aki, T., Monta\~na, A., \& Hughes,
 D.H.\ 2003, 
\mnras, submitted {\tt [astro-ph/0304178]}

\bibitem{grs01}
German, G., Ross, G.~G., \& Sarkar, S.\ 2001,
\npb, 608, 423

\bibitem{nu}
Gonzalez-Garcia, M.~C., \& Nir, Y.\ 2003,
\rmp, 75, 345

\bibitem[Grainge \ETAL 2003]{vsa} 
Grainge, K. \ETAL (VSA collab.)\ 2003, 
\mnras, submitted {\tt [astro-ph/0212495]}

\bibitem[Halverson, N.~W. \ETAL 2002]{DASI} 
Halverson, N.~W. \ETAL (DASI collab.)\ 2002,
\apj, 568, 38

\bibitem{maxima} 
Hanany, S. \ETAL (MAXIMA collab.)\ 2000,  
\apj, 545, L5

\bibitem{2dfnew}
Hawkins, E. \ETAL (2dFGRS collab.)\ 2002,
\mnras, submitted {\tt [astro-ph/0212375]}

\bibitem[Henry(1997)]{henry} 
Henry, J.~P.\ 1997, 
\apjl, 489, L1

\bibitem{cmb}
Hu, W., \& Dodelson, S.\ 2002,
\araa, 40, 171

\bibitem{jusk}
Juszkiewicz, R., Ferreira, P.~G., Feldman, H.~A., Jaffe, A.~H., \& 
 Davis, M.\ 2000,
\sci, 287, 109

\bibitem[Kinney (2001)]{fool} 
Kinney, W.~H.\ 2001, 
\prd, 63, 043001  

\bibitem{ks}
Kochanek, C.~S., \& Schechter, P.~L.\ 2003, 
{\tt [astro-ph/0306040]}

\bibitem{kogut}
Kogut, A. \ETAL (WMAP collab.) 2003\ 
\apj, submitted {\tt [astro-ph/0302213]}

\bibitem{kfbs}
Koopmans, L.~V.~E., Fassnacht, C.~D., Blandford, R.~D., \& Surpi, G.\ 2003
\apj, submitted {\tt [astro-ph/0306216]}

\bibitem[Kuo \ETAL 2002]{acbar} 
Kuo, C.~L. \ETAL (ACBAR collab.) 2002, 
{\tt [astro-ph/0212289]}

\bibitem[Leibundgut(2000)]{bl2000} 
Leibundgut, B.\ 2000, 
\aapr, 10, 179

\bibitem[Leibundgut(2001)]{bl2001} 
Leibundgut, B.\ 2001, 
\araa, 39, 67

\bibitem[Lewis \ETAL (2000)]{camb} 
Lewis, A., Challinor, A., \& Lasenby, A.\ 2000, 
\apj, 538, 473

\bibitem{Linde}
Linde, A.D.\ 1990,
``Particle Physics \& Inflationary Cosmology'' (Chur: Harwood Academic)

\bibitem[Lineweaver \ETAL 1997]{lb3}
Lineweaver, C., Barbosa, D., Blanchard, A., \& Bartlett, J.\ 1997, 
\aap, 322, 365 
        
\bibitem[Lineweaver \& Barbosa(1998)]{1998ApJ...496..624L} 
Lineweaver, C.~H.~\& Barbosa, D.\ 1998, \apj, 496, 624 

\bibitem[Lyth \& Riotto 1999]{lr99} 
Lyth, D.~H., \& Riotto, A\ 1999, 
\physrep, 314, 1

\bibitem{ls96}
Lyth, D.~H., \& Stewart, E.~D.\ 1996,
\prd, 53, 1784

\bibitem[Maddox \ETAL 1990]{apm} 
Maddox, S.~J., Efstathiou, G., Sutherland, W.~J., \& Loveday, J.\ 1990, 
\mnras, 242, 43P 

\bibitem[Markevitch 1998]{markevitch} 
Markevitch, M.\ 1998, 
\apj, 504, 27

\bibitem{mould} 
Mould, J.~R. \ETAL (HST Key Project collab.)\ 2000,
\apj, 529, 786

\bibitem[Mukherjee \& Wang 2003]{wang} 
Mukherjee, P., \&  Wang, Y.\ 2003, 
\apj, submitted {\tt [astro-ph/0303211]}

\bibitem[Myers \ETAL \ 2003]{myers}
Myers, A.~D., Shanks, T., Outram, P.~J., \& Wolfendale, A.~W.\ 2003,
\mnras, submitted {\tt [astro-ph/0306180]}

\bibitem[Netterfield \ETAL 1995]{net95} 
Netterfield, C.~B., Jarosik, N., Page, L., Wilkinson, D., \& Wollack E.\ 
 1995,
\apj, 445, L69  

\bibitem{nolta}
Nolta, M. \ETAL\ 2003,
\apj, submitted {\tt [astro-ph/0305097]}

\bibitem{parodi} 
Parodi, B.~R., Saha, A., Sandage, A., \& Tammann, G.~A.\ 2000, 
\apj, 540, 634

\bibitem[Peacock(1997)]{peac} 
Peacock, J.~A.\ 1997, 
\mnras, 284, 885 

\bibitem[Pearson \ETAL 2002]{cbi} 
Pearson, T.J. \ETAL (CBI collab.)\ 2002, 
\apj, submitted {\tt [astro-ph/0205388]}

\bibitem{pr02} 
Peebles,  P.~J.~E., \& Ratra, B.\ 2002, 
\rmp, 75, 599

\bibitem{peiri03}
Peiris, H.~V. \ETAL (WMAP collab.) 2003\ 
\apj, submitted {\tt [astro-ph/0302225]}

\bibitem[Percival \ETAL (2001)]{2df} 
Percival, W.~J. \ETAL (2dFGRS collab.)\ 2001, 
\mnras, 327, 1297 

\bibitem{scp} 
Perlmutter, S. \ETAL (SCP collab.)\ 1999, 
\apj, 517, 565 

\bibitem{pg}
Primack, J., \& Gross, M.\ 2000,
in ``Current Aspects of Neutrino Physics'', Caldwell, D.O. (ed.),
Springer (2000) p.287.

\bibitem{rp88} 
Ratra, B. \& Peebles, P.\ 1988, 
\prd 37, 3406 

\bibitem{sz}
Reese, E.~D., Carlstrom, J.~E., Joy, M., Mohr, J.~J., Grego, L., \& 
 Holzapfel, W.L.\ 2002,
\apj, 81, 53

\bibitem[Reichart \ETAL (1999)]{reich} 
Reichart, D.~E. \ETAL\ 1999,
\apj, 518, 521 

\bibitem[Reiprich, \& B{\"o}hringer 2002]{reiprich} 
Reiprich, T.~H., \& B{\"o}hringer, H.\ 2002, 
\apj, 567, 716

\bibitem{hzs} 
Riess, A.~G. \ETAL (High-z Supernova Search Team)\ 2000, 
\aj, 116, 1009

\bibitem{rs95}
Ross, G.~G., \& Sarkar, S.\ 1996,
\npb, 461, 597

\bibitem[Roussel, Sadat \& Blanchard 2000]{roussel} 
Roussel, H., Sadat, R., \& Blanchard, A.\ 2000,
\aap, 361, 429  

\bibitem[Rowan-Robinson (2002)]{rowan} 
Rowan-Robinson, M.\ 2002, 
\mnras, 332, 352  

\bibitem{rowan2} 
Rowan-Robinson, M.\ 2003,
in preparation

\bibitem[Ruhl \ETAL 2002]{boom2} 
Ruhl, J.~E. \ETAL (BOOMERanG collab.)\ 2002, 
{\tt [astro-ph/0212229]}

\bibitem[Sadat \& Blanchard 2001]{sadat2} 
Sadat, R., \& Blanchard, A.\ 2001, 
\aap, 371, 19 

\bibitem[Sadat, Blanchard \& Oukbir 1998]{sadat} 
Sadat, R., Blanchard, A., \& Oukbir, J.\ 1998, 
\aap, 329, 21  

\bibitem{sar96}
Sarkar, S.\ 1996,
\rpp, 59, 1493

\bibitem[Scott \ETAL 1996]{sco96} 
Scott, P.~F. \ETAL (VSA collab.)\ 1996, 
\apj, 461, L1

\bibitem[Seljak 2002]{seljak} 
Seljak, U.\ 2002, 
\mnras, 337, 769  

\bibitem[Sheth, \& Tormen 1999]{ST} 
Sheth, R.~K. \& Tormen, G.\ 1999, 
\mnras, 308, 119 

\bibitem{st87}
Silk, J., \& Turner, M.~S.\ 1987,
\prd, 35, 419

\bibitem{cobe1}
Smoot G.~F. \ETAL (COBE collab.)\ 1992,
\apj 396, L1

\bibitem[Spergel \ETAL 2003]{spergel} 
Spergel D. N., \ETAL (WMAP collab.)\ 2003, 
\apj, submitted {\tt [astro-ph/0302209]}

\bibitem{sdss}
Szapudi, I. \ETAL (SDSS collab.)\ 2002,
\apj, 570, 75

\bibitem[Tegmark \ETAL 2003]{teg03} 
Tegmark, M., de Oliveira-Costa, A., \& Hamilton A.\ 2003, 
{\tt [astro-ph/0302496]}

\bibitem[Tegmark, Hamilton \& Xu 2002]{2df2} 
Tegmark, M., Hamilton, A.~J.~S., \& Xu, Y.\ 2002, 
\mnras, 335, 887 

\bibitem[Uzan 2003]{uzan03a} 
Uzan J.-P., Kirchner, U., \& Ellis, G.~F.~R.\ 2003, 
{\tt [astro-ph/0302597]} 

\bibitem{shear}
Van Waerbeke, L., Tereno, I., Mellier, Y., \& Bernardeau, F.\ 2002,
{\tt [astro-ph/0212150]}

\bibitem[Vauclair \ETAL 2003]{vauclair} 
Vauclair S., \ETAL 2003, 
in preparation

\bibitem{bias}
Verde, L. \ETAL (2dFGRS collab.)\ 2002,
\mnras, 335, 432 

\bibitem[Verde \ETAL 2003]{verde} 
Verde, L. \ETAL (WMAP collab.)\ 2003, 
\apj, submitted {\tt [astro-ph/0302218]}

\bibitem[Viana, \& Liddle 1999]{viana} 
Viana, P.~T.~P.~\& Liddle, A.~R.\ 1999, 
\mnras, 303, 535  

\bibitem{wein}
Weinheimer, C.\ 2002,
{\tt [hep-ex/0210050]}

\bibitem{wett}
Wetterich, C.\ 1988,
\npb, 302, 668

\bibitem{wssd95}
White, M.~J., Scott, D., Silk, J., \& Davis, M.\ 1995,
\mnras, 276, L69

\bibitem[White \ETAL 1993]{white}  
White, S.~D.~M., Navarro, J.~F., Evrard, A.~E., \& Frenk, C.~S.\ 1993, 
\nat, 366, 429

\bibitem{wit}
Witten, E.\ 2001,
{\tt [hep-th/0106109]}


\bibitem{zar} 
Zaritsky, D.\ 1999, 
\aj, 118, 2824

\end{thebibliography}
\end{document}